\documentclass{bmcart}

\usepackage{amsthm,amsmath}
\usepackage[utf8]{inputenc} 
\usepackage{graphicx}

\startlocaldefs
\endlocaldefs

\begin{document}

\begin{frontmatter}

\begin{fmbox}
\dochead{Research}


\title{Quantum Simulation of Rindler transformations}


\author[
   addressref={1},                   
   email={csl@iff.csic.es}   
]{\inits{C}\fnm{Carlos} \snm{Sab{\'{i}}n}}


\address[id=1]{
  \orgname{Instituto de F\'isica Fundamental, CSIC, 
   Madrid, Spain}, 
  \street{Serrano 113-bis},                     %
  \postcode{28006}                                
  \city{Madrid},                              
  \cny{Spain}                                    
}



\end{fmbox}


\begin{abstractbox}

\begin{abstract} 
We show how to implement a Rindler transformation of coordinates with an embedded quantum simulator. A suitable mapping allows to realise the unphysical operation in the simulated dynamics by implementing a quantum gate on an enlarged quantum system. This enhances the versatility of embedded quantum simulators by extending the possible in-situ changes of reference frames to the non-inertial realm.
\end{abstract}




\end{abstractbox}
%

\end{frontmatter}



\section*{Introduction}

The original conception of a quantum simulator \cite{quantsimrev} as a device that implements an involved quantum dynamics in a more amenable quantum system has given rise to a wealth of experiments in a wide variety of quantum platforms such as cold atoms, trapped ions, superconducting circuits and photonics networks \cite{quantsimcold,quantsimsc,quantsimtrap,quantsimphot}. In parallel, an alternate approach to quantum simulations has been developed in the last years allowing to observe in the laboratory physical phenomena beyond experimental reach, such as {\it Zitterbewegung} \cite{zitt1, zitt2} or Klein paradox \cite{klein1, klein2,potwopot}. Moreover, along this vein a quantum simulator can also be used to implement an artificial dynamics that has only been conceived theoretically, such as the Majorana equation \cite{majorana, majorana2} or even the action of a mathematical transformation such as charge conjugation \cite{majorana, majorana2}, time and spatial parity operations \cite{qsimnonc, experiment1, qsimnonciones} or switching among particle statistics \cite{switchingembed}. These transformations are unphysical, in the sense that there is no physical operation that directly implements them in the laboratory. 

Coordinate transformations are another instance of unphysical operation. Indeed, the instantaneous application of a coordinate transformation would violate the laws of relativity. However, in \cite{qsimnonc, qsimnonciones} it is shown that {\it linear} coordinate transformations -including Galileo boosts- can be implemented in an {\it embedded quantum simulator}. An embedded quantum simulator is a device which enables the realisation of unphysical operations by encoding the dynamics of the simulated system in a larger Hilbert space. The instantaneous action of a suitable physical operation in the enlarged Hilbert space corresponds to the action of the unphysical operation in the simulated space. This is a multiplatform notion that has been initially explored in ion traps \cite{qsimnonciones, experiment1}.

In this work we show how to implement a Rindler transformation in an embedded quantum simulator. Rindler transformations are coordinate transformations suitable to describe the relation between an inertial and a uniformly accelerated observer. They are a central ingredient of quantum field theory \cite{qftbook} and are also at the core of the analysis of relativistic effects in quantum technologies \cite{rqireview}. Since they are highly non-linear operations, Rindler transformations are not included in the linear scenario considered in \cite{qsimnonc}. However, we show here that they are indeed implementable in an embedded quantum simulator. Our results allow to explore not only the non-relativistic regime -where we recover a Galileo boost, as expected- but the ultra-relativistic case as well, where the Rindler observer has been accelerated to velocities close to the speed of the light. In this way, we enhance the versatility of quantum simulation by widening the range of possible {\it in-situ} changes of reference frames that can be realised in the laboratory, paving the way to the simulation of new physical phenomena in a single-particle relativistic quantum-mechanical framework, ranging from twin-paradox scenarios to the acceleration generated by a black hole.

\section*{Rindler transformations}

Let us now provide a detailed presentation of our results. Throughout the manuscript we will use natural units $\hbar=c=1$. We start by a brief description of Rindler transformations \cite{rindlerbook}. 

A uniformly accelerated observer in 1+1 D is well described by Rindler coordinates $(\tau,\chi)$, 
where $\chi$ is related to the uniform proper velocity $a$:
\begin{equation}\label{eq:chiara}
\chi=\frac{1}{a}.
\end{equation}

Using Eq. (\ref{eq:chiara}), the transformation of coordinates between an inertial Minkowski observer and the non-inertial Rindler observer moving with uniform proper acceleration is:
\begin{eqnarray}\label{eq:rind mink}
t=&\chi\sinh(\frac{\tau}{\chi})\nonumber\\
x=&\chi\cosh(\frac{\tau}{\chi}).
\end{eqnarray}
The accelerated observer follows a trajectory:
\begin{equation}\label{eq:chi}
\chi=\sqrt{x^2-t^2}= \operatorname{constant},
\end{equation}
and $\tau=\chi\operatorname{arctanh}(t/x)$ is her proper time. Note that $\tau=\operatorname{arctanh}(t/x)/a$. 

It is convenient to introduce the rapidity $\phi=a\,\tau$, which is related to the velocity $v$ through $v=\operatorname{tanh}(\phi)$. Both magnitudes characterise the boost with respect to an inertial system that coincides with the Rindler frame at $t=\tau=0$. Unlike the standard Lorentz boosts in special relativity, in this case the velocity is space and time dependent. However, in the limit $a\rightarrow\infty$, $v=c=1$ and the Rindler observer follows a photon trajectory $x=t$.

These coordinates do not only describe the case of uniform mechanical acceleration in flat spacetime but also an observer trying to keep a fixed position in the presence of a Schwarzschild black hole, as expected due to the principle of equivalence. In both cases, an interesting feature is the existence of an horizon, splitting the spacetime in two causally disconnected regions. 

\section*{Rindler transformations in an embedded quantum simulator}

 Rindler transformations are highly non-linear and thus do not belong to the certain class of non-linear transformations discussed in \cite{qsimnonc}. However, we will see below that the embedding techniques developed in \cite{qsimnonc} can be extended to include this case.
In order to see this,  we consider now a basic dynamics governed by the equation $i\partial_t \psi = -i\partial_x\psi$. This is a $1+1$ Dirac equation for a massless particle where, for simplicity, we have traced out the internal degrees of freedom. 
Let us split the wave function $\psi$ and an arbitrary operator $\theta$  as:
\begin{eqnarray}
\psi(x,t) &=& \frac{1}{2}\bigg\{\big[ \psi(x,t) + \psi(\chi, \tau)\big]  + \big[ \psi(x,t) - \psi(\chi, \tau)\big]\bigg\}, \nonumber\\
\theta(x,t) &=& \frac{1}{2}\bigg\{\big[ \theta(x,t) + \theta(\chi, \tau)\big]  + \big[ \theta(x,t) - \theta(\chi, \tau)\big]\bigg\}.\nonumber\\ 
\end{eqnarray}
Correspondingly, for the particular case $\theta = \partial_{t,x}$, the time and spatial derivative operators are $\partial_{t,x} = \frac{1}{2}[\partial_{t,x} + \partial_{\tau,\chi}] + \frac{1}{2}[\partial_{t,x}-\partial_{\tau,\chi}]$.
 With these mappings, we can write the dynamical equation, $i\partial_t\psi~=~-i\partial_x\psi$, in terms of its even $(e)$ and odd $(o)$  components as follows,
\begin{equation}\label{1+1eo}
i(\partial^{e}_t + \partial^{o}_t)(\psi^{e} + \psi^{o}) = -i(\partial^{e}_x + \partial^{o}_x)(\psi^{e} + \psi^{o}),
\end{equation}
where $\psi^{e, o} = \frac{1}{2}\big[ \psi(x,t) \pm \psi(\chi, \tau)\big]$,~$\partial_{t,x}^{e, o} =  \frac{1}{2}[\partial_{t,x}~\pm~\partial_{\tau,\chi}]$.

Now, we define a spinor $\Psi(x, t)$  in the {\it enlarged space}, according to $\Psi = (\psi^e, \psi^o)^T$,  where $T$ is the transpose operation. The spinor $\Psi$ is related to $\psi$ through the expression $\psi(x, t) = (1, 1)\Psi$. Moreover, since $\psi(\chi,\tau)=\psi^e-\psi^o$, then the spinor $\Psi'$ in the enlarged space corresponding to $\psi(\chi, \tau)$, is just $\sigma_z \Psi$, i.e., $\psi(\chi, \tau) = (1, 1)\sigma_z\Psi(x, t)$. This means that a physical action like $\sigma_z$, acting on the enlarged space, gives rise to a physically-forbidden action -an instantaneous Rindler transformation- on the wave function in the simulated space.

The dynamical equation for $\Psi(x, t)$ can be obtained from Eq.~(\ref{1+1eo}) separating its even and odd components,  giving rise to
\begin{equation}\label{matrixeq}
i\left(\begin{array}{cc}
\partial_t^e& \partial_t^o\\
\partial_t^o& \partial_t^e\\
\end{array}\right)\left(\begin{array}{c}
\psi^e\\
\psi^o
\end{array}\right) = -i \left(\begin{array}{cc}
\partial_x^e& \partial_x^o\\
\partial_x^o& \partial_x^e\\
\end{array}\right)\left(\begin{array}{c}
\psi^e\\
\psi^o
\end{array}\right).
\end{equation}
We can write $\partial_{t, x}^{e, o}$ in terms of $\partial_t$ and $\partial_x$ as follows,
\begin{eqnarray}
\partial^{e, o}_t &=& \frac{1}{2} \big[\partial_t \pm \partial_{\tau}  \big]  = \frac{1}{2} \big[\partial_t \pm (\sinh{\frac{\tau}{\chi}}\partial_{x}+\cosh{\frac{\tau}{\chi}}\partial_{t})  \big]\nonumber\\ &=& \frac{1}{2}\bigg[\partial_t \pm  (\sqrt{a^2x^2-1}\,\partial_{x}+a\,x\partial_{t}) \bigg] ,
\end{eqnarray}
where in the last step we have used that $\chi=1/a$,
and also
 \begin{eqnarray}
\partial^{e, o}_x &=& \frac{1}{2} \big[\partial_x \pm \partial_{\chi}  \big]  =
\frac{1}{2} \big[\partial_x \pm\big(\cosh{\frac{\tau}{\chi}}-\frac{\tau}{\chi}\sinh{\frac{\tau}{\chi}})\partial_{x}+\nonumber\\ 
& & (\sinh{\frac{\tau}{\chi}}-\frac{\tau}{\chi}\cosh{\frac{\tau}{\chi}})\partial_{t}\big)  \big] \,= \nonumber \\ & & \frac{1}{2}\bigg[\partial_x \pm   ((a\,x-\sqrt{a^2x^2-1}\operatorname{arctanh}(\frac{\sqrt{a^2x^2-1}}{ax}))\partial_{x}+\nonumber \\ & & (\sqrt{a^2x^2-1}-a x\operatorname{arctanh}(\frac{\sqrt{a^2x^2-1}}{ax}))\partial_{t}) \bigg] .
\end{eqnarray}

We can substitute these expressions in Eq.~(\ref{matrixeq}) in order to obtain a Schr\"odinger equation for $\Psi$. After some algebra, we can write it in this way:
\begin{equation}\label{Psit}
i\partial_t \Psi = -i \big[f(x) I + g(x) \sigma_x \big]\partial_x\Psi,
\end{equation}
with 
\begin{eqnarray}\label{eq:embed}
f(x)&=&\frac{(a\,x+\sqrt{a^2x^2-1})(1-\frac{\operatorname{arctanh}(\frac{\sqrt{a^2x^2-1}}{a\,x})}{2})}{a\,x+\sqrt{a^2x^2-1}-a\,x\,\operatorname{arctanh}(\frac{\sqrt{a^2x^2-1}}{a\,x})}\nonumber\\
g(x)&=&\frac{(\sqrt{a^2x^2-1}-a\,x)\operatorname{arctanh}(\frac{\sqrt{a^2x^2-1}}{a\,x})}{2 (a\,x(1-\operatorname{arctanh}(\frac{\sqrt{a^2x^2-1}}{a\,x})+\sqrt{a^2x^2-1})}.
\end{eqnarray}
These equations are valid as long as the denominator is non-zero, that is, $a\,x+\sqrt{a^2x^2-1}\neq a\,x\,\operatorname{arctanh}(\frac{\sqrt{a^2x^2-1}}{a\,x})$.This is due to the fact that, in order to obtain Eq. (\ref{Psit}), we need to manipulate a equation of the form $iA\partial_t=-iB\partial_x$, where $A$ and $B$ are $2\times 2$ matrices. Therefore, we need to invert $A$, which is only invertible if the above condition is met. Otherwise, the entries of $A$ are all $1$, and the dynamics cannot be described by a Dirac-like dynamics.

\begin{figure}[h!]
\includegraphics[width=0.95\linewidth]{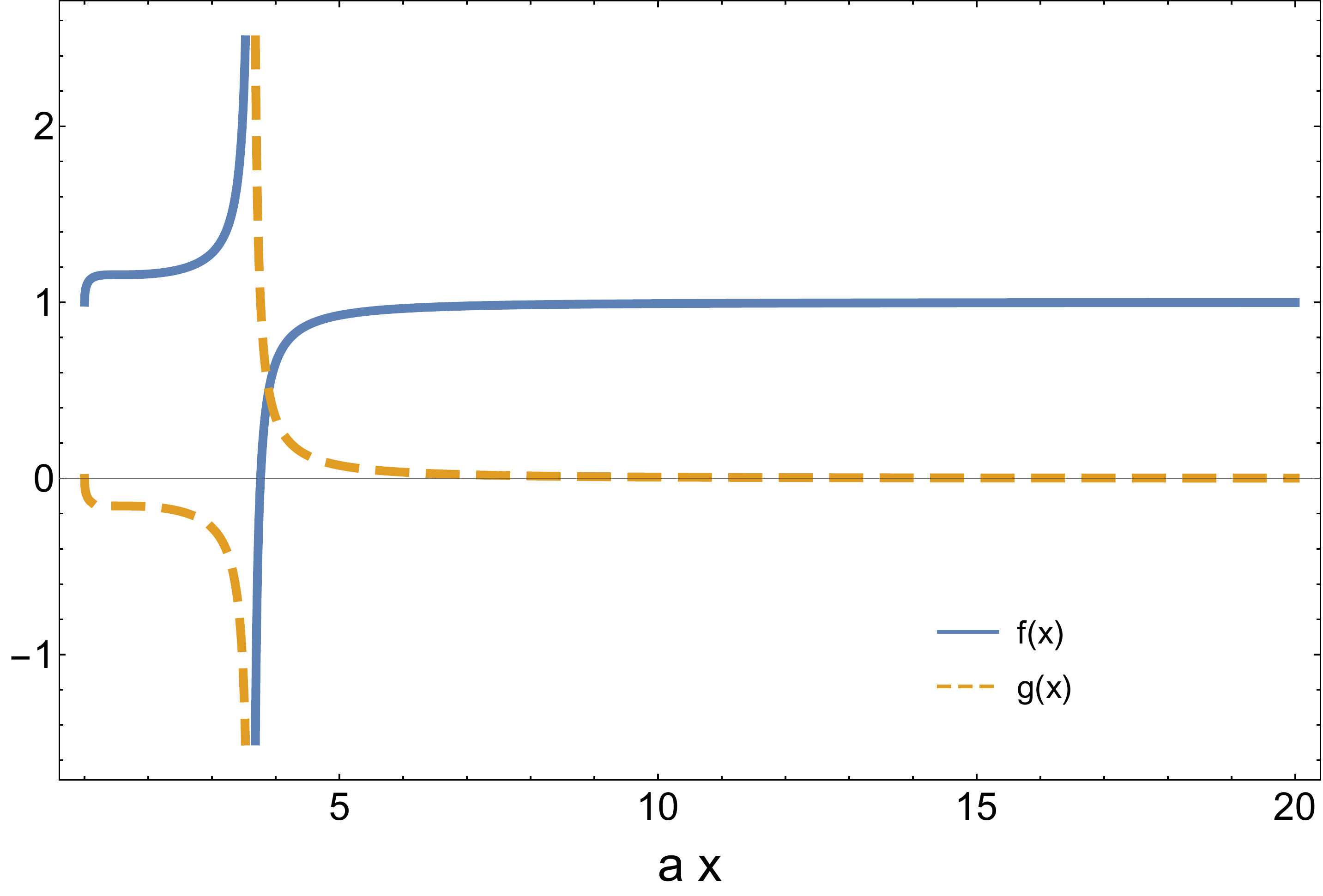}
  \caption{\csentence{Hamiltonian}
  Functions $f(x)$ and $g(x)$ in Eq. (\ref{eq:embed}), ranging from $a x=1$, which corresponds to $v=0$ and $a x=20$, which corresponds to $v=0.998749$ (in  $c=1$ units). There is a singularity at the point where the denominators are 0, which corresponds to a point where the dynamics cannot be described by a Dirac equation. We see as well the linear region in the non-relativistic regime for very small $a\,x$ $(v\simeq 0)$ and the transition to the region $v=1$ $(a\,x\rightarrow \infty)$, where the dynamics is equivalent to $v=0$, due to Lorentz-invariance.} \label{Fig1}
      \end{figure}

Using that $a\,x=\cosh[\operatorname{arctanh}(v)]$ we can rewrite Eq. (\ref{eq:embed}) in terms of the velocity boost only. In Fig. (\ref{Fig1}), we see the behavior of $f(x)$ and $g(x)$ ranging from $a x=1$ ($v=0$) and $a x=20$, ($v=0.998749\simeq c$)

We are now able to analyse two interesting regimes.  We consider first the non-relativistic regime where $v\simeq\phi<<1$, and then $a\,x\simeq1+v^2/2$. By considering this limit in Eq. (\ref{eq:embed}), we recover, as expected the embedded dynamics of a Galileo boost \cite{qsimnonc}:
\begin{eqnarray}\label{eq:gal}
i\partial_t \Psi = -i \big[(1+\frac{v}{2}) I -\frac{v}{2} \sigma_x \big]\partial_x\Psi.
\end{eqnarray}
Notice however that in this case $v$ is spacetime dependent, unlike in standard Galileo boosts.

Now we consider the opposite regime, that is, an ultra relativistic observer $v=1-\delta$, where $0<\delta<<1$. In this case, $a\,x=\cosh[\operatorname{arctanh}(1-\delta)]$. Expanding in $\delta$, we obtain:
\begin{eqnarray}\label{eq:ultrar}
i\partial_t \Psi = -i \big[\big(1+f(\delta)\big) I -f(\delta) \sigma_x \big]\partial_x\Psi.
\end{eqnarray}
where 
\begin{equation}\label{eq:fultrar}
f(\delta)=-\frac{\delta}{2(1+\frac{4}{\log[\frac{\delta}{2}]})}. 
\end{equation}
Note that is restricted to values of $\delta$ that are sufficiently far from $\log[\frac{\delta}{2}]=-4$, which corresponds to $a\,x+\sqrt{a^2x^2-1}= a\,x\,\operatorname{arctanh}(\frac{\sqrt{a^2x^2-1}}{a\,x})$, which is the singular point described above. Under this additional condition, we can write:
\begin{equation}
f(\delta)\simeq-\frac{\delta}{2}\simeq-\frac{1}{4\,a^2\,t^2}.
\end{equation}
Notice that in the limit $\delta=0$ ($v=c=1$) we obtain the same trivial dynamics as in $v=0$. This is because in the case $v=c$, the Rindler transformation becomes a standard Lorentz time-independent boost, that is, the transformed reference frame is inertial. Accordingly, the coordinate transformation does not change the Lorentz-invariant Dirac dynamics in the simulated space.

In order to determine the dynamics associated with Eq.~(\ref{Psit}), one just has to define the initial condition for $\Psi$, i.e., $\Psi(x,0) = \frac{1}{2} [ \psi(x,0) + \psi \big(\chi(x,0), \tau(x,0) \big) , \psi(x,0) - \psi \big(\chi(x,0), \tau(x,0) \big) ]^T$. 
Notice that in this case $ \psi \big(\chi(x,0), \tau(x,0) \big)=\psi(x,0)$ and therefore the dynamics is determined by the knowledge of the initial wave function only.

The techniques of \cite{qsimnonc} for relating observables in the enlarged and simulated spaces are also applicable here.
We can obtain any expectation value of either the inertial or Rindler wave functions through observables in the enlarged space as follows:
\begin{eqnarray}
\langle O \rangle_{\psi(x, t)} = \langle\psi| O |\psi\rangle &=& \langle\Psi|\left(\begin{array}{c}
1\\
1\end{array}\right) O  \left(\begin{array}{cc} 1  &,1\end{array}\right)|\Psi\rangle\nonumber\\ &=& \langle\Psi| (I + \sigma_x)\otimes O |\Psi\rangle,\\
\langle O \rangle_{\psi(\chi, \tau)} = \langle\psi'| O |\psi'\rangle &=& \langle\Psi|\sigma_z\left(\begin{array}{c}
1\\
1\end{array}\right) O  \left(\begin{array}{cc} 1  &,1\end{array}\right)\sigma_z|\Psi\rangle\nonumber\\ &=& \langle\Psi| (I - \sigma_x)\otimes O |\Psi\rangle,
\end{eqnarray}
where we use $\langle x|\psi\rangle = \psi(x, t)$, $\langle \chi|\psi'\rangle = \psi(\chi, \tau)$, and $\langle x|\Psi\rangle=\Psi(x, t)$. We are also able to analyse correlations between $\psi(x, t)$ and $\psi(\chi, \tau)$ out of the dynamics in the enlarged space only:
\begin{eqnarray}
\langle O \rangle_{\psi(x, t), \psi(\chi, \tau)}  = \langle\psi| O |\psi'\rangle &=& \langle\Psi|\left(\begin{array}{c}
1\\
1\end{array}\right) O  \left(\begin{array}{cc} 1  &,1\end{array}\right)\sigma_z|\Psi\rangle\nonumber\\ &=& \langle\Psi| (\sigma_z - i \sigma_y)\otimes O |\Psi\rangle.
\end{eqnarray}

For instance, this would allow to reveal the existence of twin-paradox time dilation effects through quantum measurements \cite{twinpa} as well as the degradation of correlations between an inertial and an accelerated observer close to a black hole horizon \cite{alice}.This is so because $\psi(\chi,\tau)$ would be the natural description of the spinor $\psi$ under uniform acceleration. Thus, the experiments would be straightforward in a two-particle Dirac simulator. One particle would remain inertial -thus subject to the standard Dirac Hamiltonian- while the other would undergo a period of simulated uniform acceleration by means of the implementation of the Rindler transformation -in the black hole case- or a trajectory with several acceleration and deceleration steps -several Rindler transformations- and several inertial steps, in order to simulate a twin-paradox trajectory. Then, a comparison of the time coordinates of the two particles would allow to measure a simulated relativistic time dilation and the measurement of the two-particle correlations would allow to detect a degradation of correlations due to acceleration or gravity. This degradation might be linked with the black-hole information problem, since it suggests that quantum information cannot be a solution for the information loss. However, a deeper analysis of this problem would require the simulation of a more complete theory of quantum gravity. While quantum field theory phenomenology is out of reach in this single-particle experiment, one main advantage would be the possibility of simulating higher values of the acceleration. 

\section*{Possible experimental implementations}

A Dirac-like equation, such as Eq. (\ref{eq:ultrar}) can be implemented in several quantum platforms with current technology. The trapped-ion setup envisaged in \cite{qsimnonciones} for the implementation of Eq. (\ref{eq:gal}) can also accommodate Eq. (\ref{eq:ultrar}) with a modification of the experimental parameters, thus widening the range of the simulator from non-relativistic to ultra-relativistic physics. In particular, a suitable time-dependent frequency trap would produce the required Lamb-Dicke parameter for the simulation. More specifically, in \cite{qsimnonciones} it has been shown that a trapped-ion quantum simulator with realistic experimental parameters \cite{zitt2} can achieve good fidelities in the case of Eq. (\ref{eq:gal}), for simulated values of $(c+v/2)$ set by a frequency range from $\nu/100$ to $\nu/25$ -where $\nu$ is the frequency of the trap. These velocities are comparable to the simulated value of $c$ in \cite{zitt2}, which is given by $\nu/20$. In $c=1$ units, this means that we can achieve as well good fidelities in the simulation of Eq. (\ref{Psit}) with experimental parameters, as long as $f(x)$ is close to 1. In Fig \ref{Fig1}, we see that this is indeed the case almost everywhere, except for a small region near the singularity. 

In superconducting circuit architectures, it has been suggested that Dirac dynamics are also achievable, both in the absence and presence of external potentials \cite{julen}. The proposed setup consists of one superconducting qubit interacting with a single mode of a superconducting resonator, with suitable classical drivings. With the addition of the techniques developed in this paper, it can be used as an alternative approach to the analysis of acceleration \cite{teleportation, joel} and gravity \cite{rmp} in superconducting circuits. Indeed, the wide tunability of parameters that can be achieved in superconducting architectures could be exploited to obtain the dynamics in Eq. (\ref{eq:ultrar}). Letting alone the conceptual differences and the benefits of the embedding approach, our framework would allow to analyse extreme ultra-relativistic regimes and is experimentally simpler than current proposals. In \cite{teleportation, joel} the acceleration is achieved through the ultrafast variation of magnetic fluxes threading SQUIDs, and is restricted to a much more modest regime of velocities. The proposed simulation schemes for effective spacetime metrics \cite{rmp} would require a SQUID array embedded along a transmission line resonator \cite{sorin} and a suitable electromagnetic pulse travelling at the speed of light along the transmission line. Of course, our simulations would be restricted to single-particle relativistic dynamics.

\section*{Conclusion}

In summary, we have devised an embedded quantum simulator for Rindler transformation of coordinates. This unphysical mathematical operation can be mapped to a physical observable in an enlarged Hilbert space, whose action is equivalent to the change of coordinates in the simulated space. Thus, we are able to analyse expectation values of observables of both the untransformed and transformed wavefunctions as well as correlations among them with measurements on the enlarged system only. This paves the way to the analysis of extreme accelerations and black hole horizons in quantum platforms such as trapped ions and superconducting circuits, within a single-particle relativistic quantum-mechanical approach.

\begin{backmatter}

\section*{Declarations}
\section*{Availability of data and material}
Not applicable.
\section*{Competing interests}
  The authors declare that they have no competing interests.
\section*{Funding}
Financial support by Fundaci{\' o}n General CSIC (Programa ComFuturo) is acknowledged.  

\section*{Author's contributions}
 Not applicable.

\section*{Acknowledgements}
 Not applicable.
\section*{Author's information}
Not applicable.

\end{backmatter}
\end{document}